\documentclass{article} 
\usepackage{nips15submit_e,times}
\usepackage{hyperref}
\usepackage{url}
\hypersetup{colorlinks=false,pdfborder=0 0 1,citebordercolor={0 1 0}}\usepackage{amsmath, amsthm, amssymb}
\usepackage{graphicx}
\usepackage{caption}
\usepackage{subcaption}
\title{Matching Community Structure Across Online Social Networks}

\author{
Lin Li\\
Human Language Technology Group \\
MIT Lincoln Laboratory\\
Lexington, MA 01740\\
\texttt{lin.li@ll.mit.edu}
\And
W. M. Campbell\\
Human Language Technology Group \\
MIT Lincoln Laboratory\\
Lexington, MA 01740\\
\texttt{wcampbell@ll.mit.edu}
}

\nipsfinalcopy 

\begin{document}
\renewcommand{\thefootnote}{*} \footnotetext{This work was sponsored
  by the Defense Advanced Research Projects Agency under Air Force
  Contract FA8721-05-C-0002. Opinions, interpretations, conclusions,
  and recommendations are those of the authors and are not necessarily
  endorsed by the United States Government.}
\maketitle
\begin{abstract}
%
%
The discovery of community structure in networks is a problem of
considerable interest in recent years. In online social networks,
often times, users are simultaneously involved in multiple social
media sites, some of which share common social relationships. It is of
great interest to uncover a shared community structure across these
networks. However, in reality, users typically identify themselves with
different usernames across social media sites. This
creates a great difficulty in detecting the community structure. In
this paper, we explore several approaches for community detection
across online social networks with limited knowledge of username alignment across the networks. We refer to the known alignment of usernames as seeds. We investigate strategies for seed
selection and its impact on networks with a different fraction of
overlapping vertices. The goal is to study the interplay between network
topologies and seed selection strategies, and to understand how it
affects the detected community structure. We also propose several
measures to assess the performance of community detection and use them
to measure the quality of the detected communities in both
Twitter-Twitter networks and Twitter-Instagram networks.

\end{abstract}

\section{Introduction}\label{sec:intro}
%
%
The problem of detecting communities is one of the most important
tasks when studying networks. Community detection can be viewed as
partitioning a graph into subgraphs with a higher probability of
interaction between vertices of the same subgraph than vertices of
different subgraphs. Finding community structure in networks often
provides important information about the function and organization of
the network.

In this paper, we explore several strategies for detecting a shared
community structure across online social networks (e.g., Twitter,
Instagram, etc.) with a focus on addressing two important issues that
often come up when dealing with data from multiple online social
networks.
\begin{itemize}
 \item {\it Unknown alignment of users across social networks}: In
   online social networks, users are identified by their
   usernames. They may use different usernames for various social
   media sites and multiple users may have similar usernames across
   different sites. Although there are methods such as named entity
   disambiguation that can help in identifying users from their public
   profiles, depending on the quality and consistency of the
   information provided across domains, only a small percentage of the
   mapping between usernames can be inferred with high confidence. We
   refer to these aligned users as {\it seeds}. A
   natural question is how to take advantage of this extra seed
   information for finding a shared community structure across
   online social networks.
 \item {\it Unknown fraction of shared users between social
   networks}: Online social networks have become important tools for
   sharing, discovery and networking. They differ in what service or
   function they provide---e.g., share pictures, write comments, make
   friends with common interests, etc. While many users are involved
   in multiple online social networks and they have common connections
   across networks, it is not clear how much overlap there is across
   real-world social networks and how the community structure is
   affected as the fraction of this overlap changes.
\end{itemize}

To address the above issues and set the scope of this paper, we assume
that there is an implicit community structure that describes the
social relationships between people, and each network observed from a
particular social media site provides partial information about this
underlying community structure. By combining information from multiple
social networks, one can find a shared community structure that more
accurately captures the overall social structure.

With the above assumption, community detection across social networks
can be useful for several important applications. First, it provides a
comprehensive understanding of users' social behaviors and
subsequently, it allows us to examine the similarities and differences
between communities across online social networks. Second, given that
we do not have the exact username alignment across networks, the
detected communities can be used as a tool to improve the named entity
disambiguation results by adjusting the scores for pairs of usernames
based on their community memberships, because similar usernames across
networks are more likely to be associated to the same person if they
belong to the same community. Lastly, community detection across two
social networks can be useful in solving large-scale graph matching
problem through a divide-and-conquer procedure \cite{lyzinski2015spectral}: vertices of the two
graphs are jointly clustered based on their community memberships and
then one can run any graph matching algorithm in parallel within each
cluster.

The rest of the paper is organized as
follows. Section~\ref{sec:relatedwork} presents previous work on
community detection in networks and some generalization to multilayer
networks. In Section~\ref{sec:commDetect}, we describe a range of
strategies for community detection across networks using only seeds
and explain several measures that we use to assess the performance of
the detection. In Section~\ref{sec:exp}, we will present performance
analysis for different community detection strategies on networks with
a different fraction of shared users and a varying number of
seeds. Our goal is to understand how the network structure (i.e.,
fraction of shared vertices) and seeds affect the detection of
communities across networks. Also, several seed selection strategies are
proposed and compared with a goal of improving the community detection performance.  The results
are presented using both Twitter-Twitter networks and
Twitter-Instagram networks. Conclusions and some future directions
are presented in Section~\ref{sec:conclusion}.
   

\section{Related Work}\label{sec:relatedwork}
%
%
Many methods for community detection have been proposed over the
years. Some of the most popular methods include hierarchical
clustering
\cite{girvan2002community,donetti2004detecting,radicchi2004defining},
modularity maximization
\cite{newman2004finding,guimera2004modularity,clauset2004finding,
wakita2007finding,blondel2008fast},
Infomap \cite{rosvall2007maps,rosvall2008maps}, clique percolation
\cite{palla2005uncovering}, Potts spin glass model approach
\cite{reichardt2006statistical,ronhovde2009multiresolution} and many
others. Lancichinetti and Fortunato have carried out a careful
comparative analysis of these methods against a set of benchmark
graphs \cite{lancichinetti2009community} (i.e., the LFR benchmark
\cite{lancichinetti2008benchmark}, the benchmark by Girvan and Newman
\cite{girvan2002community} and random graphs
\cite{erdds1959random}). The results of their tests show that Infomap
\cite{rosvall2010map}, which is based on a compression of information
on the network, has excellent performance. The modularity-based Louvain
algorithm \cite{blondel2008fast} and the Potts spin glass model
\cite{ronhovde2009multiresolution} also have good performance.  

More recently, the study of multilayer networks has become an important research direction in network
science~\cite{mucha2010community}. A general notion of a multilayer
network consists of a set of networks called {\it
layers}~\cite{mucha2010community,de2013mathematical} and a set of
interlayer edges that connect vertices across layers. Despite the
large volume of research on community detection in networks, only a
few approaches have been proposed for combining multiple layers for
community detection in multilayer networks. They can be generally
divided into three classes---aggregation~\cite{de2013mathematical},
linking~\cite{jeub2015local} (also referred to as classical random
walk), and relaxed random walk~\cite{de2015identifying}.  The
aggregation approach transforms a multi-layer network to a single
layer network by merging aligned vertices and then applies community detection on the resulting
graph.  The linking approach preserves the multilayer structure by
connecting aligned vertices in different layers and then performing community
detection on the final result.  A challenge in this case is extending
standard cost functions to the multilayer structure, e.g.,
modularity~\cite{mucha2010community}.  Finally, relaxed random walk
switches between intralayer and interlayer random walks where the
interlayer walk is to the neighbors of the aligned vertices.

All of the above methods assume that vertices across layers have a known 
alignment. As hinted earlier, this assumption is not true in our problem:
(1) we may not know the alignment because of different 
usernames; (2) a user may not appear in both networks.

\section{Matching Community Structure}\label{sec:commDetect}
%
%
In this paper, we focus on exploring various strategies for learning the 
community structure across online social networks, particularly when we have limited knowledge of the alignment between usernames across social networks. As 
mentioned earlier, many users are involved in multiple social networks. We model 
the interactions between users in various networks as a multilayer network 
$\mathcal{G} = \{G_1, G_2, \cdots, G_N\}$, where each layer is represented by a 
graph $G_n = (V_n, E_n)$ that describes the interactions between users in a 
particular social network. The key to detect communities across networks is to 
connect the networks via shared users (see Section~\ref{sec:relatedwork}). 
However, in practice users are identified by their usernames which are not 
unique across different networks. Hence, we have limited information about the 
alignment of users across networks and we call this information seeds.  

Without loss of generality and for ease of notation, we now restrict our 
attention to the two-layer network case, i.e., $\mathcal{G} = \{G_1, G_2\}$.  
Let $\mathcal{T} = \{\left(i_k, j_k\right)\}$ be the set of   all aligned users across $G_1$ and $G_2$, where $i_k\in V_1$ and $j_k \in V_2$. Let 
$\mathcal{S} = \{(i_s, j_s)\}$ denote a set of seeds and thus 
$\mathcal{S}\subset \mathcal{T}$. The goal is to obtain a shared community 
structure for the given multilayer network $\mathcal{G}$ using only seeds 
$\mathcal{S}$ and to understand the effect of $\mathcal{S}$ on the performance 
of the community detection. To this end, we first explain three approaches for 
multilayer community detection. Then we introduce three measures to assess the 
performance of the detection.

\subsection{Methods}\label{sec:methods}
\begin{figure}[hbt]
    \centering
    \subcaptionbox{Aggregation}{ \includegraphics[width=2.5cm]{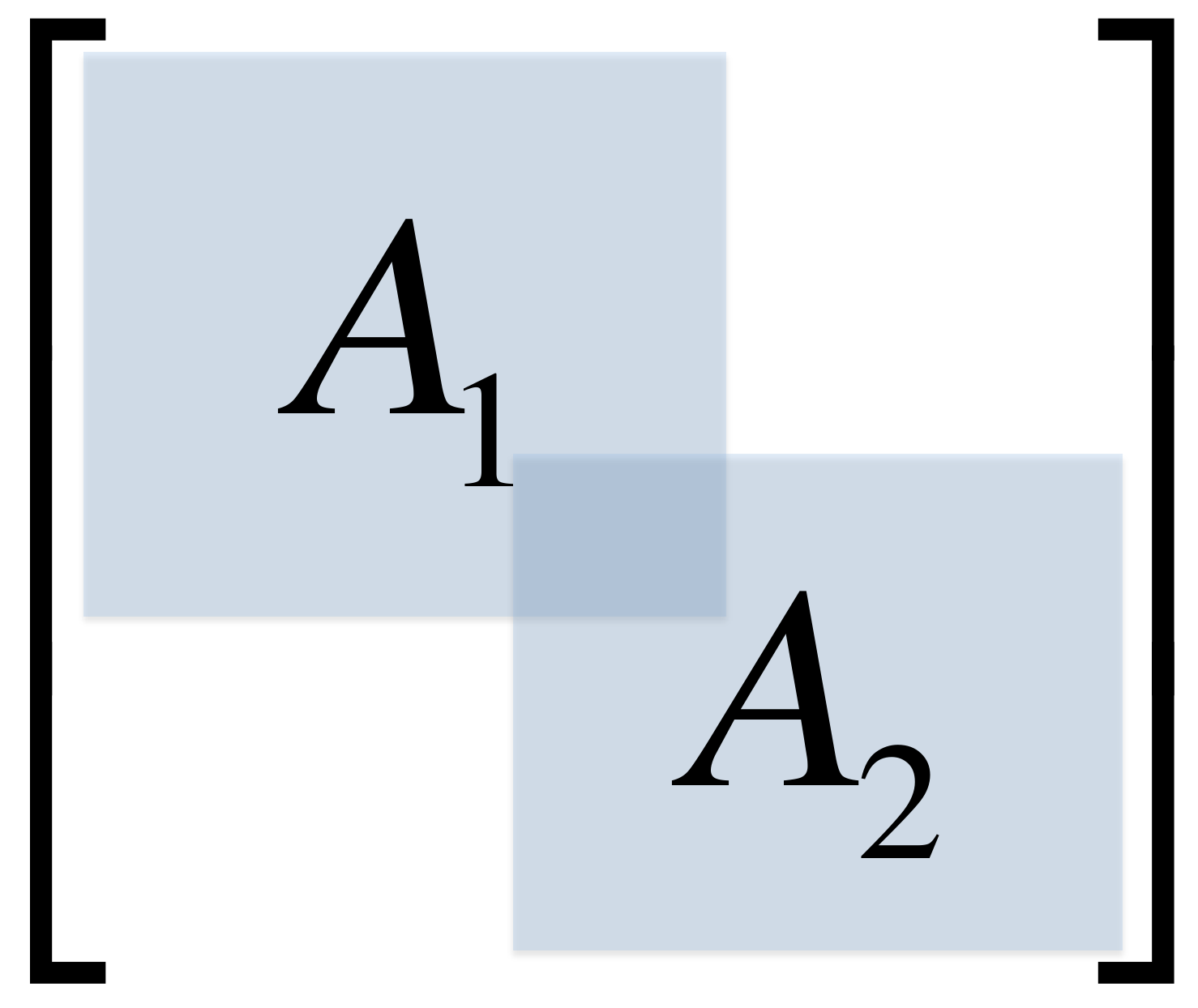}}\quad
    \subcaptionbox{Linking}{ \includegraphics[width=2.3cm]{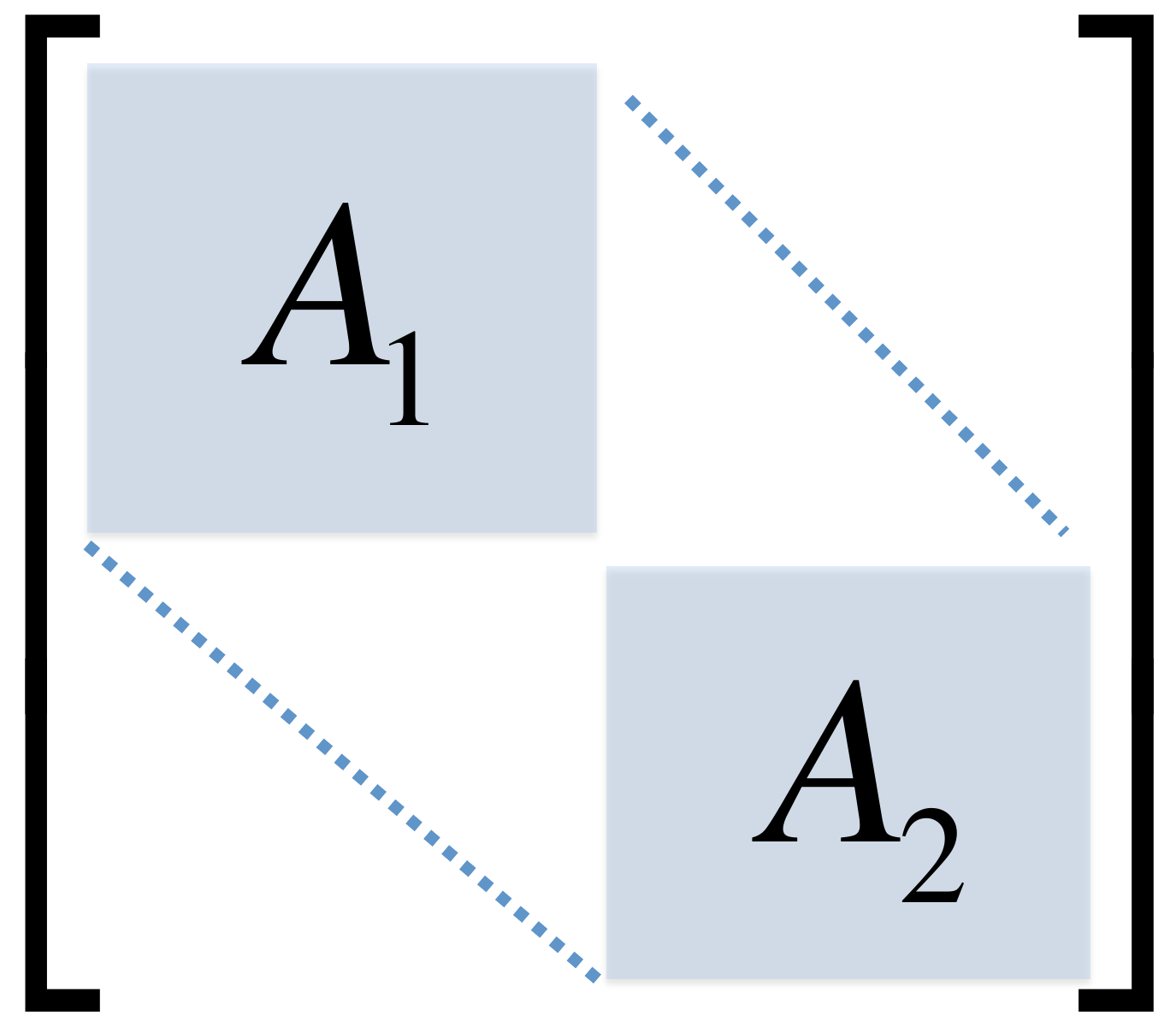}}
    \subcaptionbox{Relaxed~random~walk}{\quad\includegraphics[height=2cm]{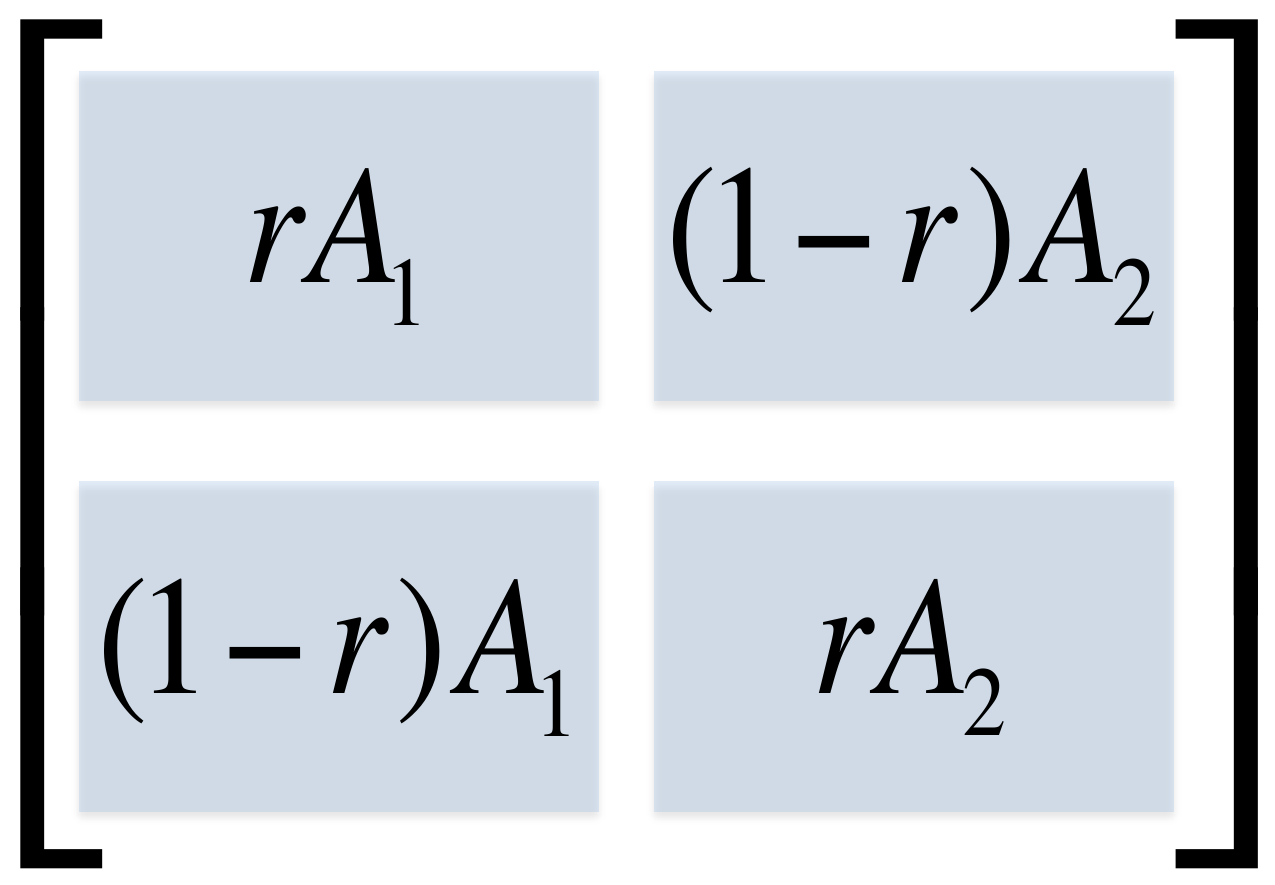}\quad}
    \caption{Supra-adjacency matrices for performing community detection in multi-layer networks}\label{fig:methods}
\end{figure}

%
%
For multilayer community detection, we combine two methodologies.
First, we focus on random walks on multilayer networks.  As mentioned
in Section~\ref{sec:relatedwork}, we apply the three approaches of
aggregation, linking, and relaxed random walk to construct a network
for analysis.  Second, we focus on Infomap~\cite{rosvall2008maps} as a
community detection algorithm.  Since Infomap is random walk based, it
is straightforward to adapt it to the various multilayer network
construction approaches. The three approaches for multilayer analysis
are shown in Figure~\ref{fig:methods} and described as follows.

{\it Aggregation approach.} This approach first aggregates the network
layers into a single (monoplex) network by merging pairs of vertices in the seed set and then applies a standard
community detection methods to this network. Moreover, if a
ground-truth community is given, then one can also compute an optimal
linear weights for combining the layers. It should also be noted that
taking the mean or the sum of the edge weights across layers can be
problematic, particularly in the case of temporal multilayer networks
because it makes a restrictive assumption on the interaction dynamics
between the vertices.  As shown in Figure~\ref{fig:methods}~(a), the
resulting supra-adjacency matrix (using the terminology
from~\cite{kivela2014multilayer}) overlaps the adjacency matrices of
the original graphs.

{\it Linking approach.} This approach combines two graph layers by
giving each layer a unique set of vertices.  The alignment between
vertices in each layer is represented by a link between the layers.
The resulting supra-adjacency matrix shown in
Figure~\ref{fig:methods}(b) contains the original adjacency matrices
on the block diagonal.  The off-diagonal blocks are diagonal matrices.
Note that in the figure, we show the case of complete connectivity
between the layers.  In the general case, the diagonals would be
sparse and positive only for seed vertices.

{\it Relaxed random walk approach.} In the relaxed walk approach, a random
walker switches between walking on one layer at some probability $r$
to switching between layers at probability $(1-r)$.  As shown in the
supra-adjacency matrix in Figure~\ref{fig:methods}(c), interlayer
walks are performed to neighbors of aligned vertices and intralayer walks
are determined by the adjacency matrix within that layer.  Note that
in the figure, we have focused on the completely connected case.  In
general, only the seed rows would be non-zero in the off-diagonal
blocks.


\subsection{Performance Measures}\label{sec:perf}
%
%

We evaluate community detection using methods that capture both
performance and community-level entity resolution accuracy.
Define the assignment of the community that vertex~$i$ belongs to as
${\rm comm}(i)$; let $C_i=\left\{j|{\rm comm}(j)=i\right\}$.  For
community detection performance evaluation, we start with a reference
community assignment $C=\left\{C_1,\dots,C_N\right\}$ and a community assignment
to test $C'=\left\{C'_1,\dots,C'_M\right\}$.  The contingency matrix,
${\cal N}$, has entries $n_{i,j}=\left|C_i\cap C'_j\right|$.
Qualitatively, we measure performance by visualizing the matrix of
Jaccard similarity indices,
\begin{equation}\label{eqn:jaccard}
J(C_i,C'_j)=\frac{\left|C_i\cap C'_j\right|}{{\left|C_i\cup C'_j\right|}}=\frac{n_{i,j}}{{\left|C_i\cup C'_j\right|}}.  
\end{equation}
Additionally, we use variation of information
(VI)~\cite{meilua2003comparing} as a quantitative distance measure of
community assignment performance,
\begin{equation}
\operatorname{VI}(C,C')=H(C|C') + H(C'|C)
\end{equation}
where $H(C|C')$ is the conditional entropy of $C$ given $C'$.  This
value can be calculated by defining $p_{i,j}=n_{i,j}/n$,
$p_i=\left|C_i\right|/n$, and $q_i=\left|C'_i\right|/n$, where $n$ is the total
number of vertices.  Then, 
\begin{equation}\label{eqn:vi}
\operatorname{VI}(C,C') = -\sum_{i,j} p_{i,j}\left[\log\frac{p_{i,j}}{p_i}+\log\frac{p_{i,j}}{q_i}\right].
\end{equation}
A desirable property of $\operatorname{VI}(\cdot,\cdot)$ is that it is a distance metric and
has an interpretable meaning in terms of information theory.

In addition to community assignment performance, we use a recall-based
measure which we call {\it oracle accuracy}.  For a set of truth
pairs, ${\cal T}$, where each truth pair is given by $(i_k,j_k)$
($i_k$ and $j_k$ correspond to the same user), we define the oracle
accuracy as
\begin{equation}\label{eqn:oa}
\operatorname{OA}(T, C')=\frac{\left|\left\{k|{\rm comm}(i_k)={\rm comm}(j_k)\right\}\right|}{|T|}.
\end{equation}


\section{Experiments}\label{sec:exp}

\subsection{Experimental Setup}\label{sec:setup}
%
%



We used geotagged Twitter and Instagram data from the Boston region
for our experiments.  For Twitter, a total number of 4,654,373
tweets were collected from 1/1/14 to 10/1/14.  For Instagram, a total
number of 3,709,479 posts and comments were collected from
12/31/2013 to 7/1/2015.

Network construction was performed by designating both users (e.g.,
@twitter) and hashtags (e.g., \#fashion) as vertices in the network.  For
Twitter, edge types corresponded to multiple categories---user-to-user
tweets, user mentions of users or hashtags, retweets,
and co-occurrence of hashtags or users.  For Instagram, edge types
corresponded to---user comments on user posts, user mentions of users,
user mentions of hashtags, and co-occurrence of users or hashtags.
For both Twitter and Instagram, the count of occurrence of various
edges types was saved in the network.  For final analysis, counts were
summed across edge types resulting in a weighted undirected graph.
More details may be found
in~\cite{campbell-baseman-greenfield:2014:SocialNLP}.


\subsection{Twitter-Twitter Networks}\label{sec:tt_exp}
%
%
To show the effect of varying the fraction of shared vertices across networks 
on 
community detection, we take the Twitter graph $G$ from which two random 
subgraphs $G_1 = (V_1, 
 E_1)$ and $G_2 = (V_2,E_2)$ are chosen such that the fraction of the shared 
vertices is given by $\{0.1, 0.5, 0.9\}$, 
where $0.1$ means that $10\%$ of the vertices belong to both $G_1$ and $G_2$.  
 
Denote $C$ as the community assignment of the original Twitter graph $G$. For 
the first experiment, we assume that the alignment of all shared vertices is
known. 
Then, we apply the methods described in 
Section \ref{sec:methods} for community detection across $G_1$ and $G_2$, and 
let $C'$ denote the resulting community assignment. Table \ref{table} displays 
the results of the three methods under the evaluation of the variation of 
information $\operatorname{VI}(C,C')$. It can be seen that the variation of 
information 
decreases (i.e., less change in the community structure) as the fraction of 
shared vertices across networks increases. Among these three methods, the 
aggregation approach has the lowest variation of information.

\begin{table}[hbt!]
\caption{The variation of information results of three different methods on 
networks with $10\%$, $50\%$ and $90\%$ shared vertices}\label{table}
\begin{center}
\begin{tabular}{l*{3}{c}}
\% Shared Vertices             & Aggregation & Linking & Relaxed RW \\
\hline
10 	& \bf{1.8301} & 1.9777 & 2.1793  \\
50      & \bf{0.6590} & 1.4523 & 1.3963  \\
90      & \bf{0.3406} & 1.3730 & 1.0793   \\
\end{tabular}
\end{center}
\end{table}

\begin{figure}[h!]
\begin{subfigure}{\linewidth}
\centering
\includegraphics[width=0.45\textwidth]{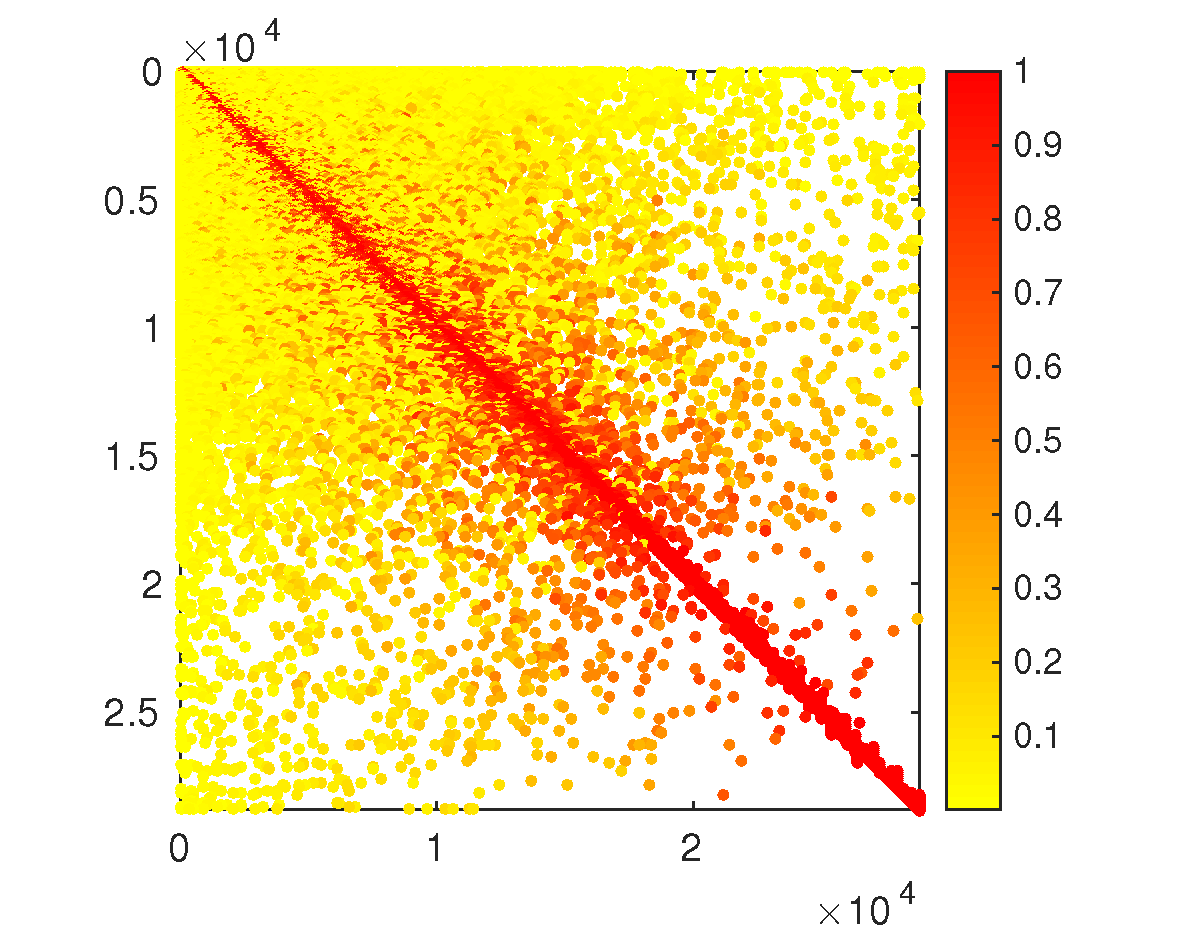}
\caption{Aggregation}
\label{fig:method1}
\end{subfigure}
\begin{subfigure}{.53\linewidth}
\includegraphics[width=1.05\textwidth]{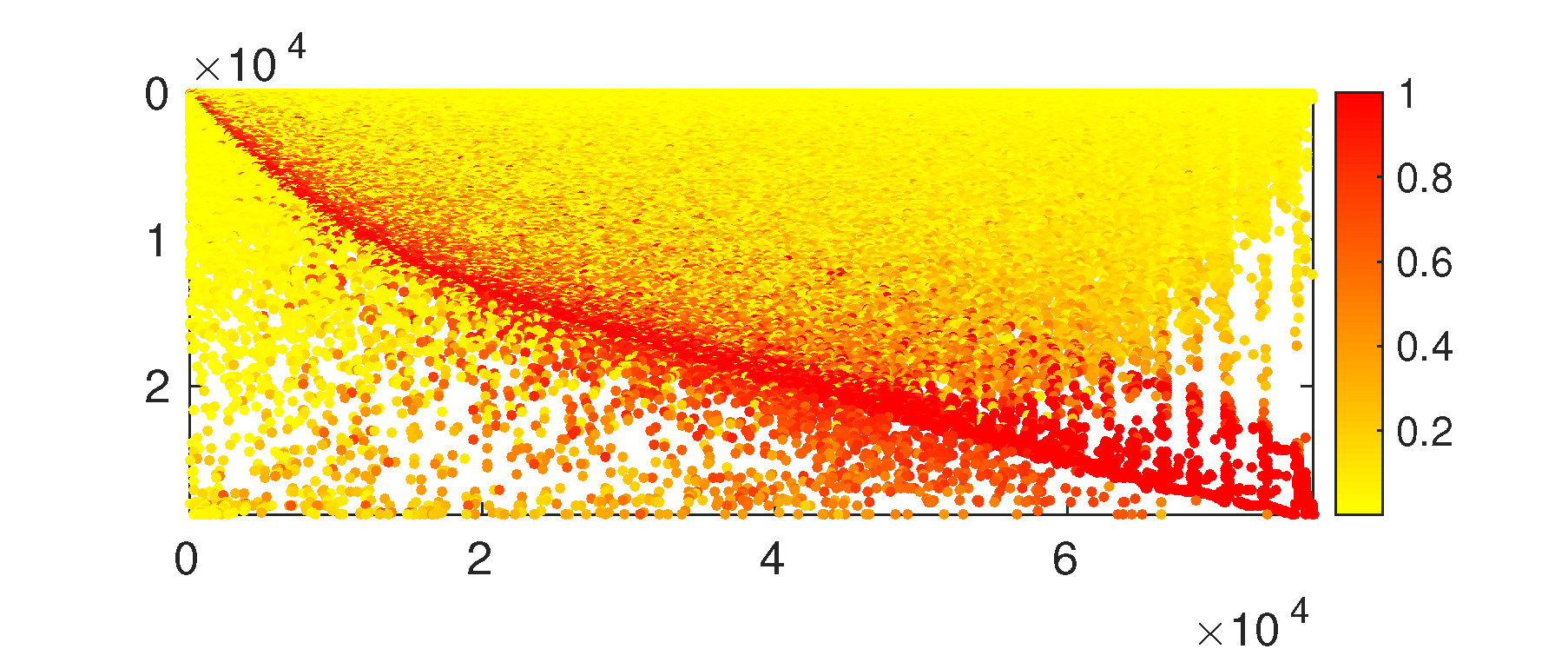}
\caption{Linking}
\label{fig:method2}
\end{subfigure}
\begin{subfigure}{.47\linewidth}
\centering
\includegraphics[width=1.05\textwidth]{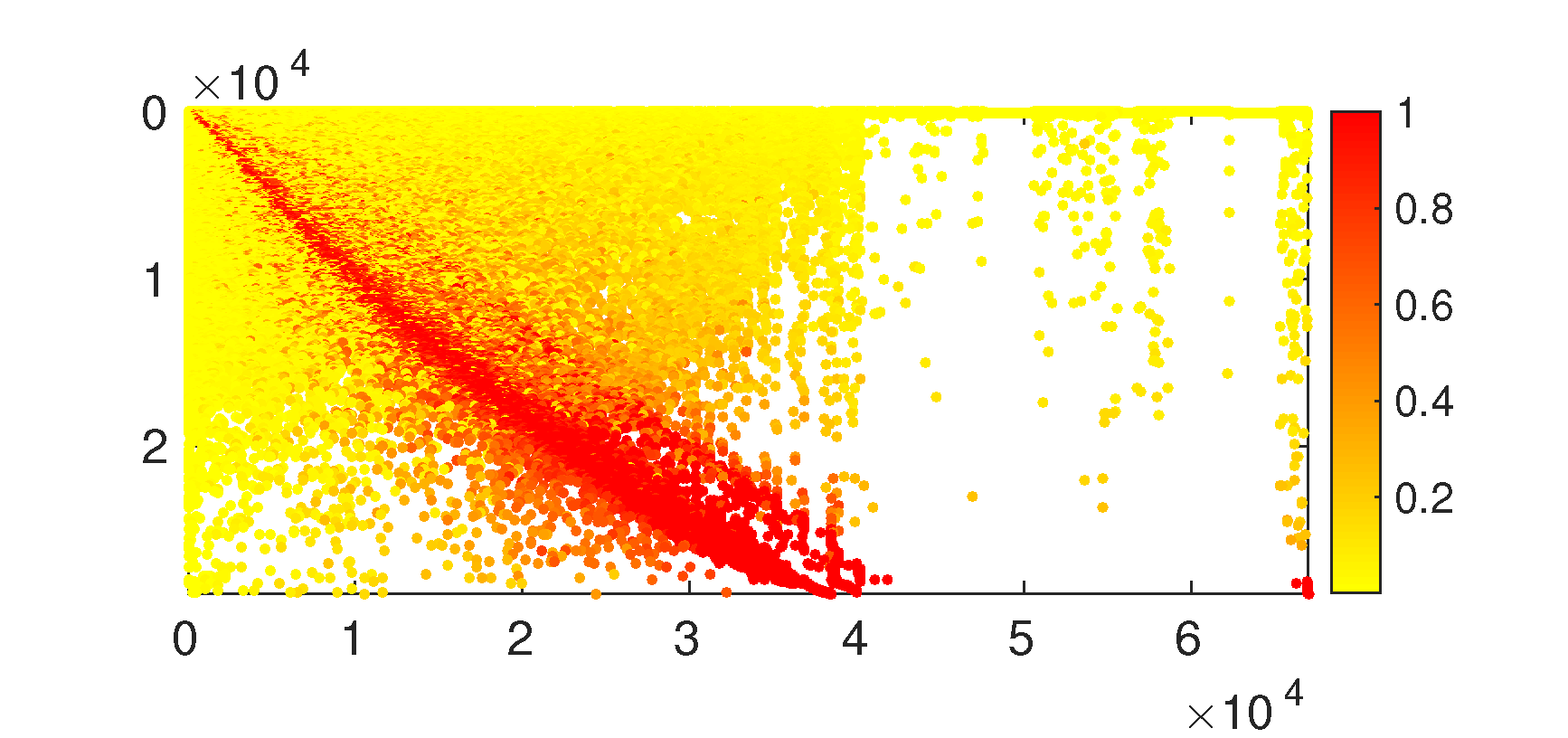}
\caption{Relaxed RW}
\label{fig:method3}
\end{subfigure}
\caption{The Jaccard matrices for networks with $90\%$ shared vertices}
\label{fig:heatmap_methods}
\end{figure}

Fig. \ref{fig:heatmap_methods} shows the Jaccard similarity matrix between $C$ 
($y$-axis) and $C'$ ($x$-axis) of 
the same experiment on networks with $90\%$ shared vertices. Communities are 
ordered by the community PageRank, i.e., index $1$ denotes the community with 
the highest 
PageRank value. Observe Fig. \ref{fig:heatmap_methods}(a) that the Jaccard 
matrix of the 
aggregation approach has large values close to the main diagonal, 
indicating that $C$ and $C'$ share similar community structure. It is to be 
noted that if the Jaccard matrix is the identity matrix, then it means that the 
corresponding communities in the reference $C$ and the detected $C'$ are 
exactly 
the same. Fig. \ref{fig:heatmap_methods}(b) and \ref{fig:heatmap_methods}(c) 
show that the resulting community assignment  
$C'$ using the linking approach and the relaxed random walk approach 
respectively, have 
more than double the number of communities than that of the reference $C$. In 
particular, from Fig. \ref{fig:heatmap_methods}(c), observe that the first few 
communities 
(i.e., communities with the highest PageRank values) in the reference $C$ break 
up into smaller communities in $C'$, many of which have low community PageRank 
values. 
This also explains why the variation of information values for these two method 
are large compared to the aggregation approach. In all remaining experiments, 
the aggregation approach is used for community detection.

\begin{figure}[hbt!]
\begin{subfigure}{.333\linewidth}
\includegraphics[width=1.05\textwidth]{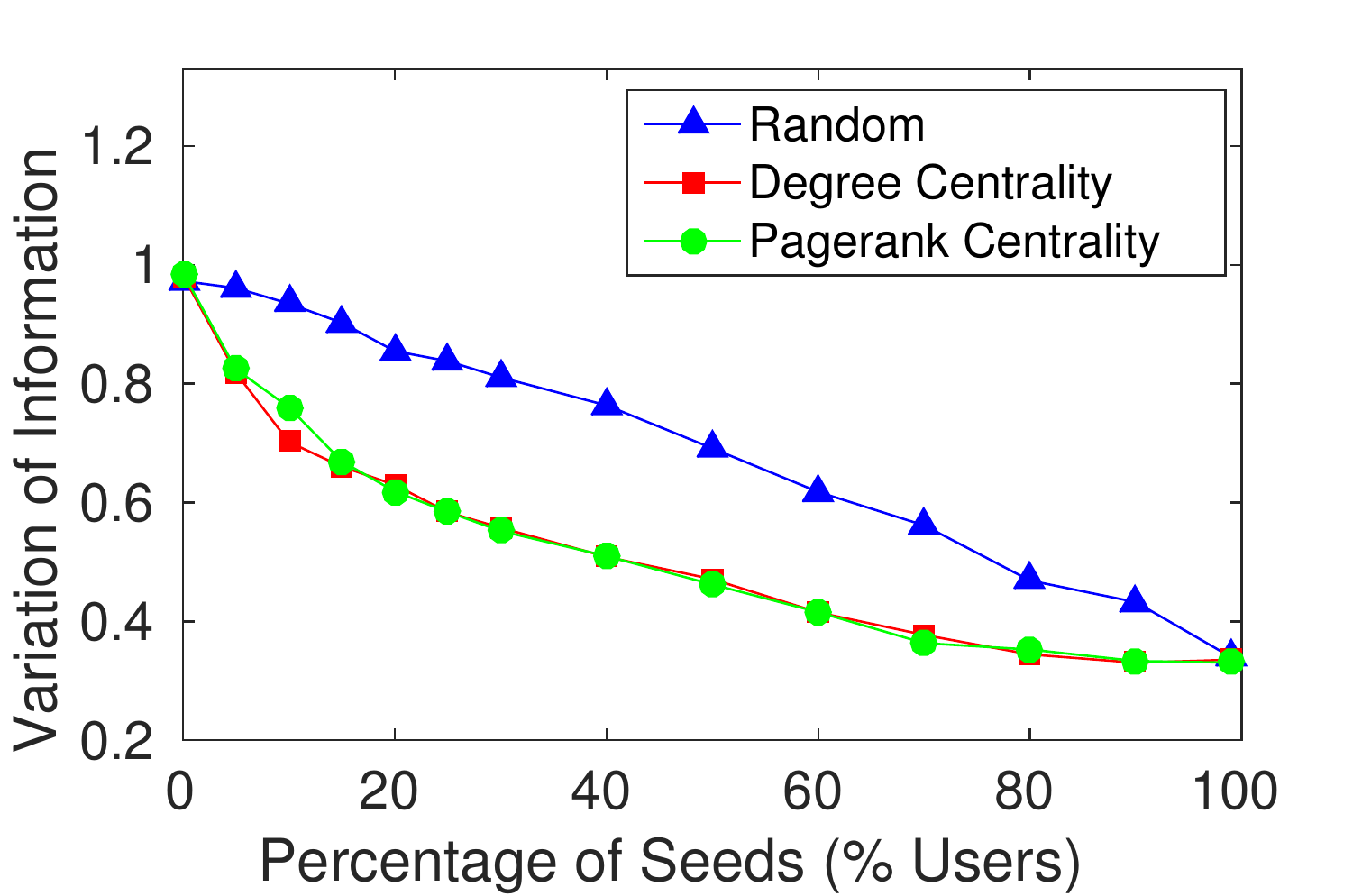}
\caption{$10 \%$ shared vertices}
\end{subfigure}
\begin{subfigure}{.333\linewidth}
\includegraphics[width=1.05\textwidth]{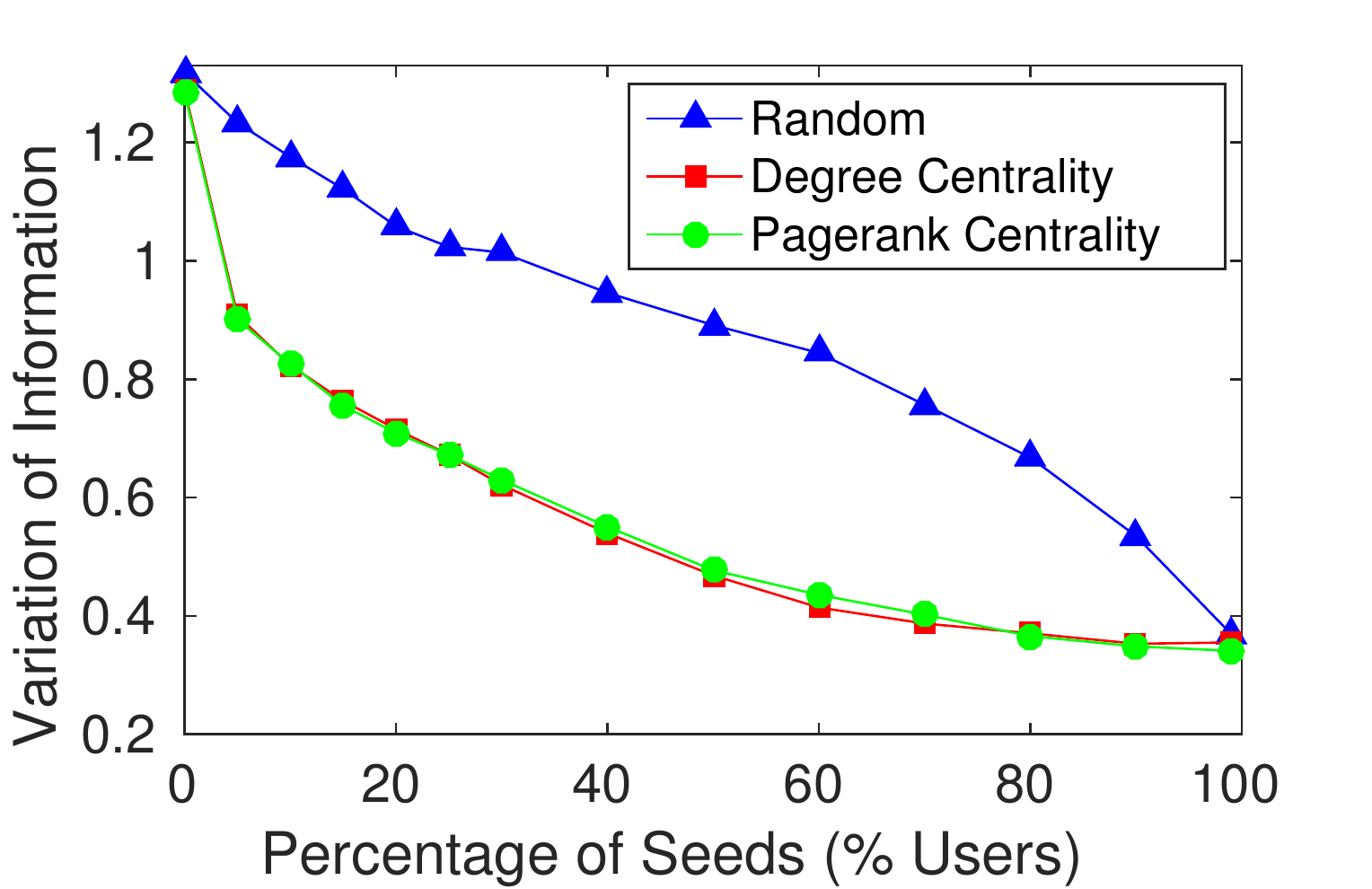}
\caption{$50 \%$ shared vertices}
\end{subfigure}
\begin{subfigure}{.333\linewidth}
\centering
\includegraphics[width=1.05\textwidth]{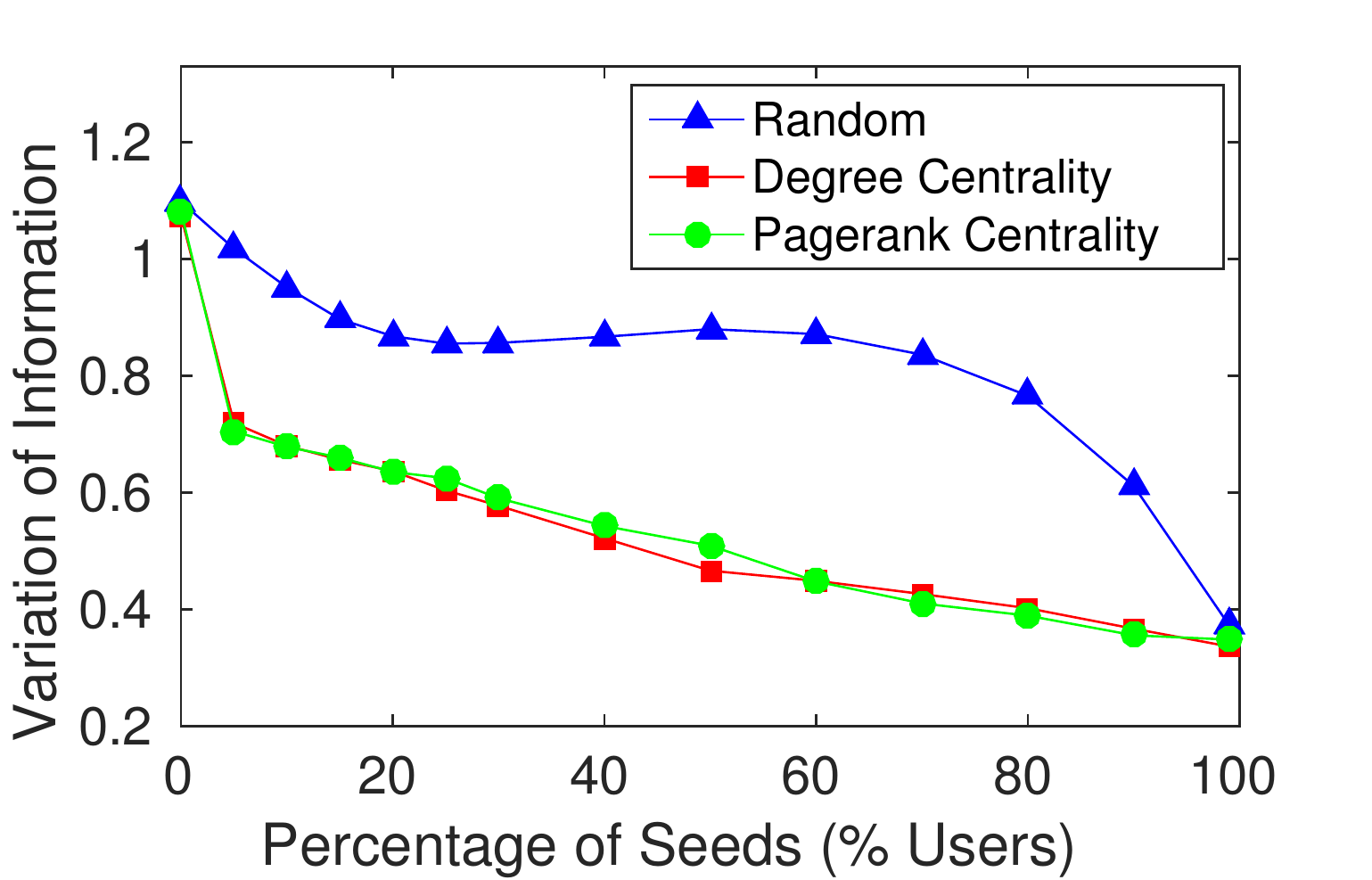}
\caption{$90 \%$ shared vertices}\label{fig:share90}
\end{subfigure}
\caption{The variation of information results as a function of the percentage 
of 
seeds}\label{seeds:vi}
\end{figure}

In the second experiment, we investigate the effect of using seeds on networks 
with a different fraction of shared vertices. Let 
$C'$ be the detected community 
assignment across $G_1$ and $G_2$ given only seeds $\mathcal{S}$. Let  $C$ be the reference 
community assignment across $G_1$ and $G_2$ using all shared users $\mathcal{T}$. The plots in Fig.~\ref{seeds:vi} 
illustrate the results using the variation of information metric 
$\operatorname{VI}(C,C')$ as a function of the percentage of seeds, i.e.,
$|\mathcal{S}|/|\mathcal{T}|$. Note that hashtags are not considered as seeds, 
and they are always aligned across networks based on exact string match. Three 
seed selection strategies are used: random, degree and PageRank. The random 
strategy randomly selects a fixed number of mappings from $\mathcal{T}$ and 
used 
them as seeds $\mathcal{S}$. The degree approach first computes the vertex 
degrees of the original Twitter graph and then selects the seeds from 
$\mathcal{T}$ that have high degree centrality. Similarly, the PageRank 
approach 
selects seeds that have high PageRank values. From  Fig.~\ref{seeds:vi}, it can 
be observed that both the degree centrality strategy and PageRank centrality 
strategy give substantially better performance than random seed selection.

\begin{figure}[t]
\begin{subfigure}{.333\linewidth}
\includegraphics[width=1.05\textwidth]{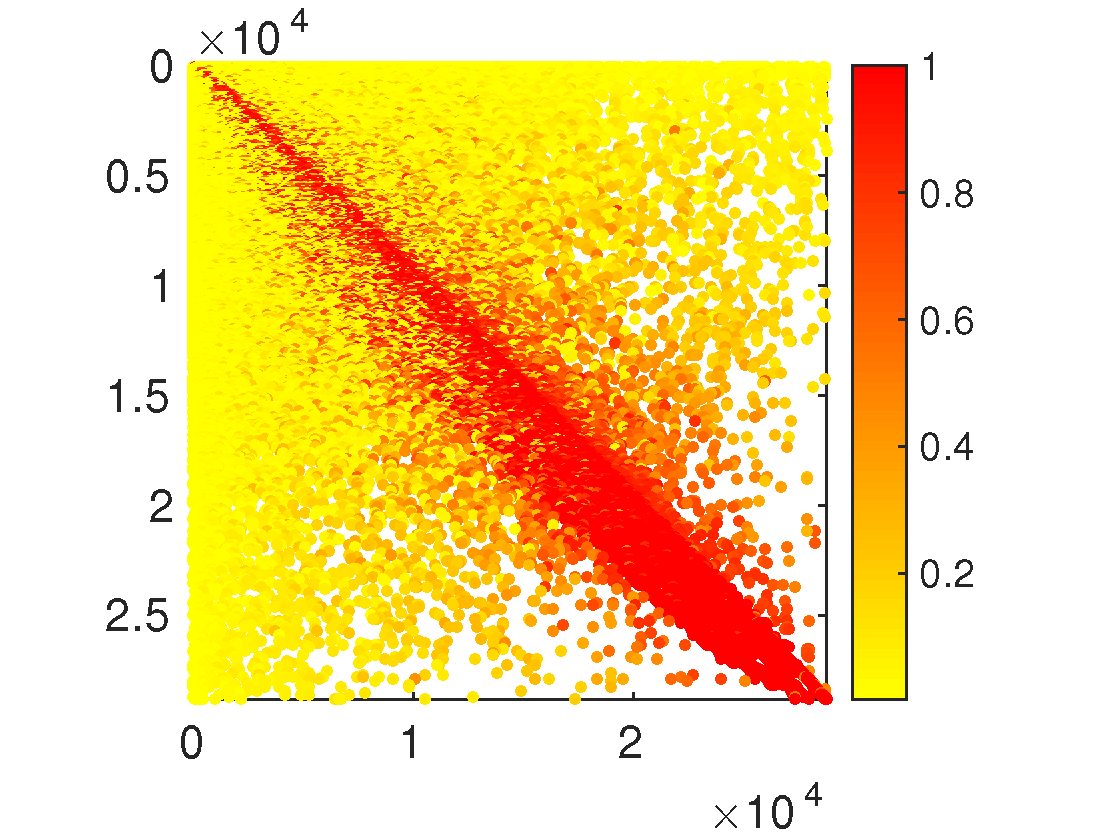}
\caption{Random}
\label{fig:random90}
\end{subfigure}
\begin{subfigure}{.333\linewidth}
\includegraphics[width=1.05\textwidth]{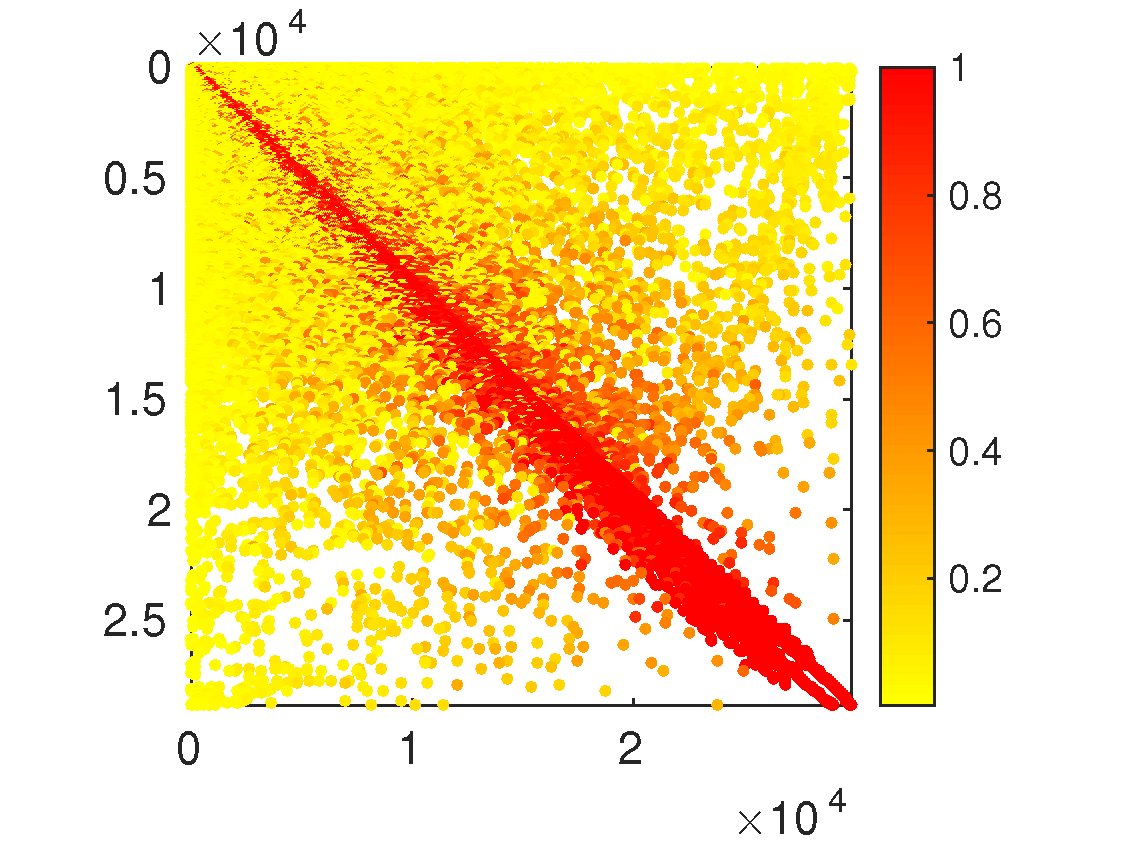}
\caption{Degree Centrality}
\label{fig:degree90}
\end{subfigure}
\begin{subfigure}{.333\linewidth}
\centering
\includegraphics[width=1.05\textwidth]{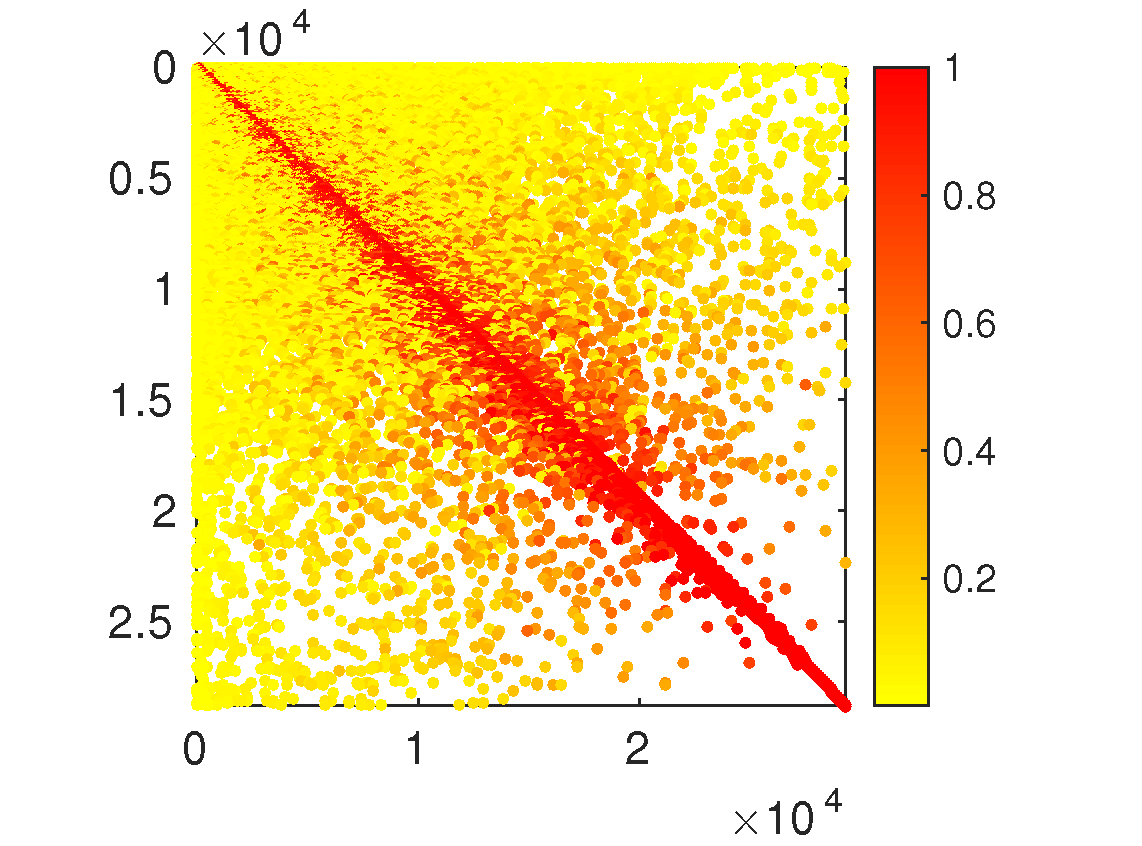}
\caption{PageRank Centrality}
\label{fig:pr90}
\end{subfigure}
\caption{The Jaccard matrices for networks with $90 \%$ shared vertices and 
using 
$90 \%$ seeds}
\label{fig:O90_S90}
\end{figure}

\begin{figure}[hbt!]
\begin{subfigure}{.333\linewidth}
\includegraphics[width=1.05\textwidth]{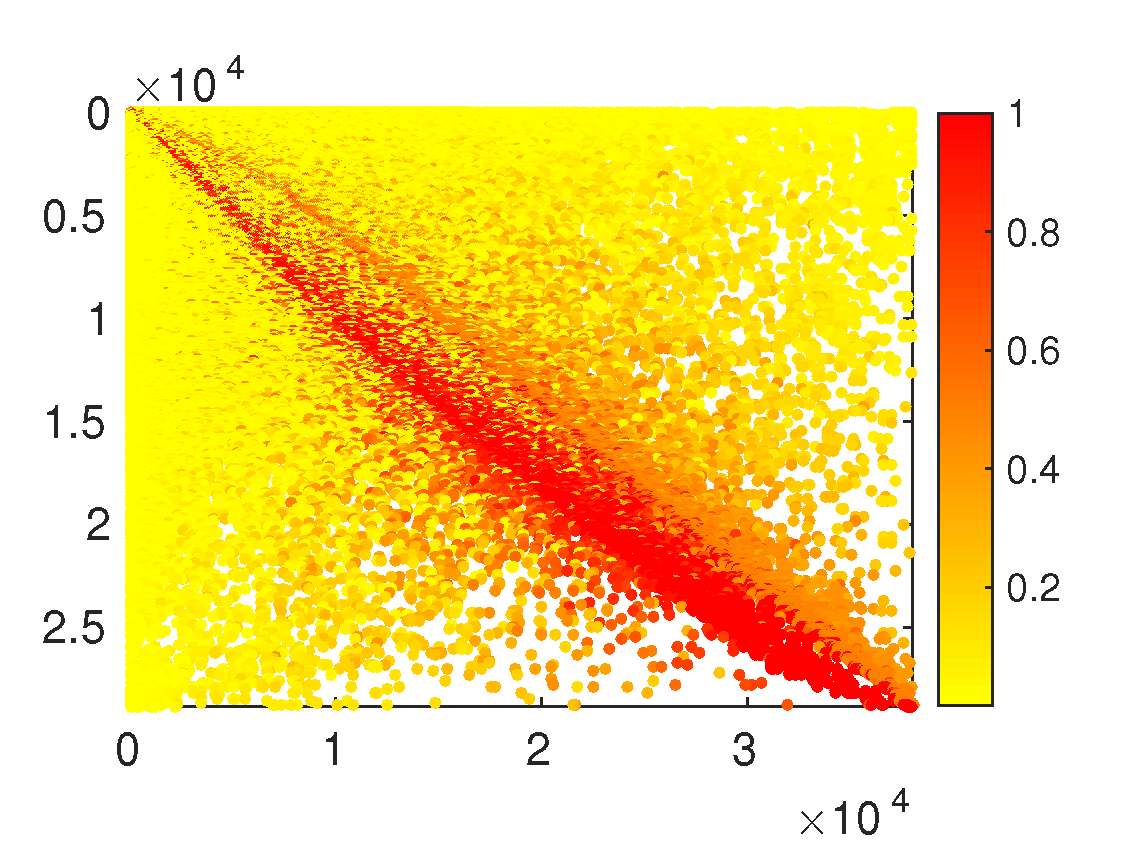}
\caption{Random}
\label{fig:random5}
\end{subfigure}
\begin{subfigure}{.333\linewidth}
\includegraphics[width=1.05\textwidth]{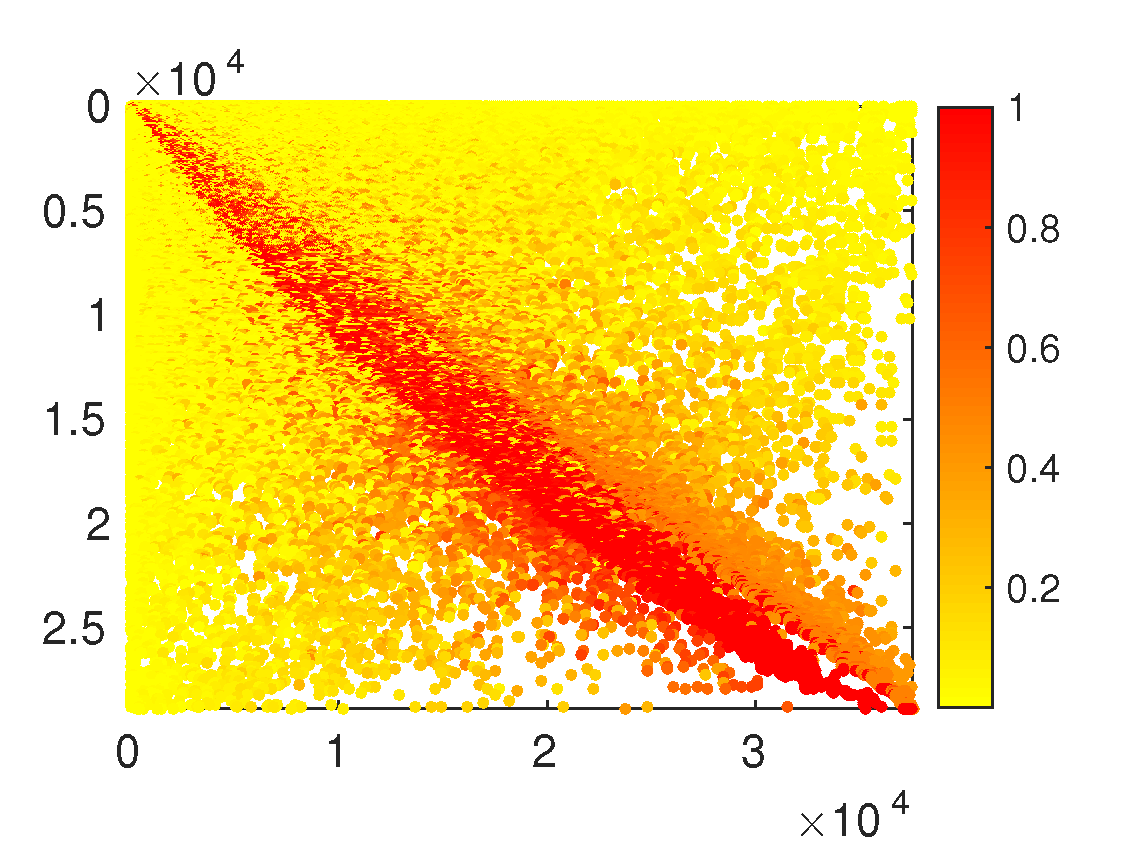}
\caption{Degree Centrality}
\label{fig:degree5}
\end{subfigure}
\begin{subfigure}{.333\linewidth}
\centering
\includegraphics[width=1.05\textwidth]{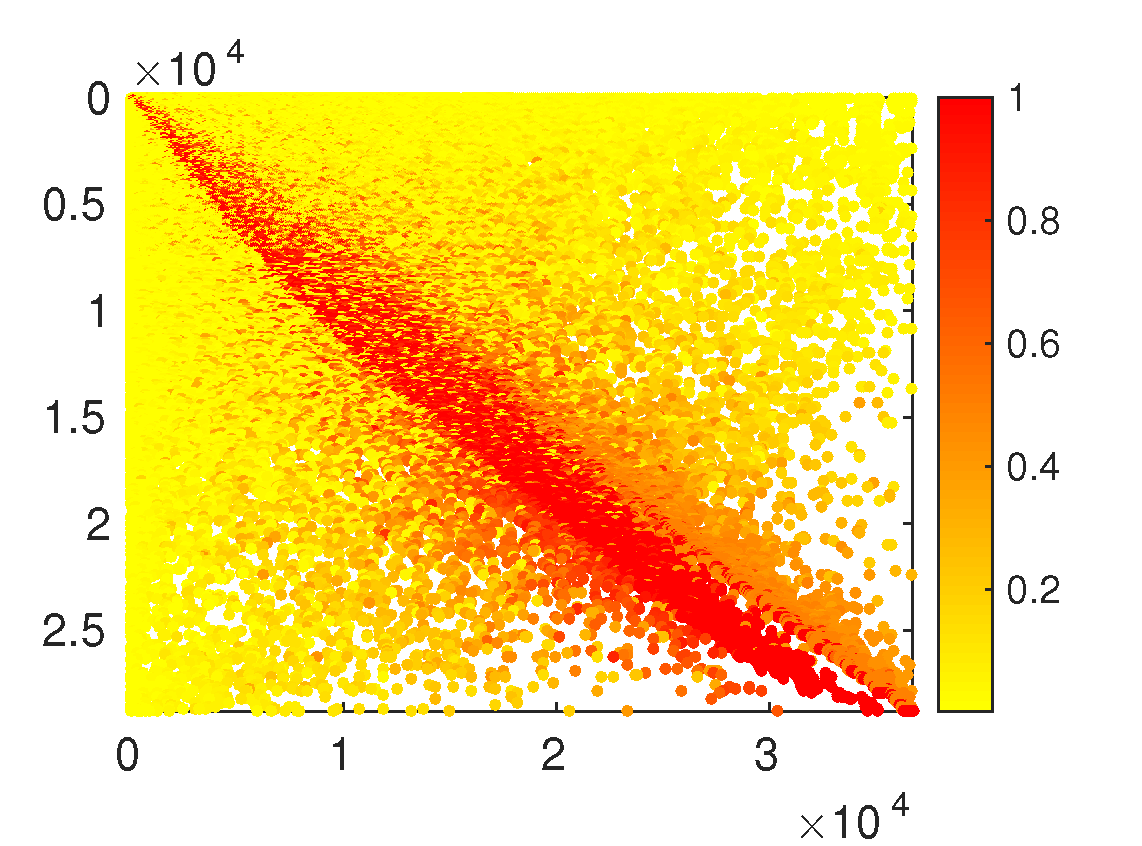}
\caption{PageRank Centrality}
\label{fig:pr5}
\end{subfigure}
\caption{The Jaccard matrices for networks with $90 \%$ shared vertices and 
using 
$5 \%$ seeds}
\label{fig:O90_S5}
\end{figure}

Fig.~\ref{fig:O90_S90} and \ref{fig:O90_S5} display the Jaccard matrices of 
different seed selection strategies. Fig.~\ref{fig:O90_S90} shows the results 
using $90\%$ seeds. Observe that the detected community structure using the 
PageRank strategy is the most similar to the reference community structure. 
Fig.~\ref{fig:O90_S5} displays the results with $5\%$ seeds. It can be observed from 
Fig.~\ref{fig:O90_S5}(a) that bifurcation occurs, breaking up one community in 
the reference $C$ into roughly two communities in the detected $C'$. This is 
likely due to the lack of useful connections between the two networks. With 
careful seed selection (i.e., degree or PageRank), the bifurcation disappears 
as 
shown in Fig.~\ref{fig:O90_S5}(b) and \ref{fig:O90_S5}(c). Although in this 
case, it is hard to see which strategy gives better performance, the Jaccard 
matrices are still useful in discovering the relationship between the reference 
$C$ and the detected $C'$ and allow us to discover local changes in the 
community structure.

\begin{figure}[t]
\begin{subfigure}{.333\linewidth}
\includegraphics[width=1.05\textwidth]{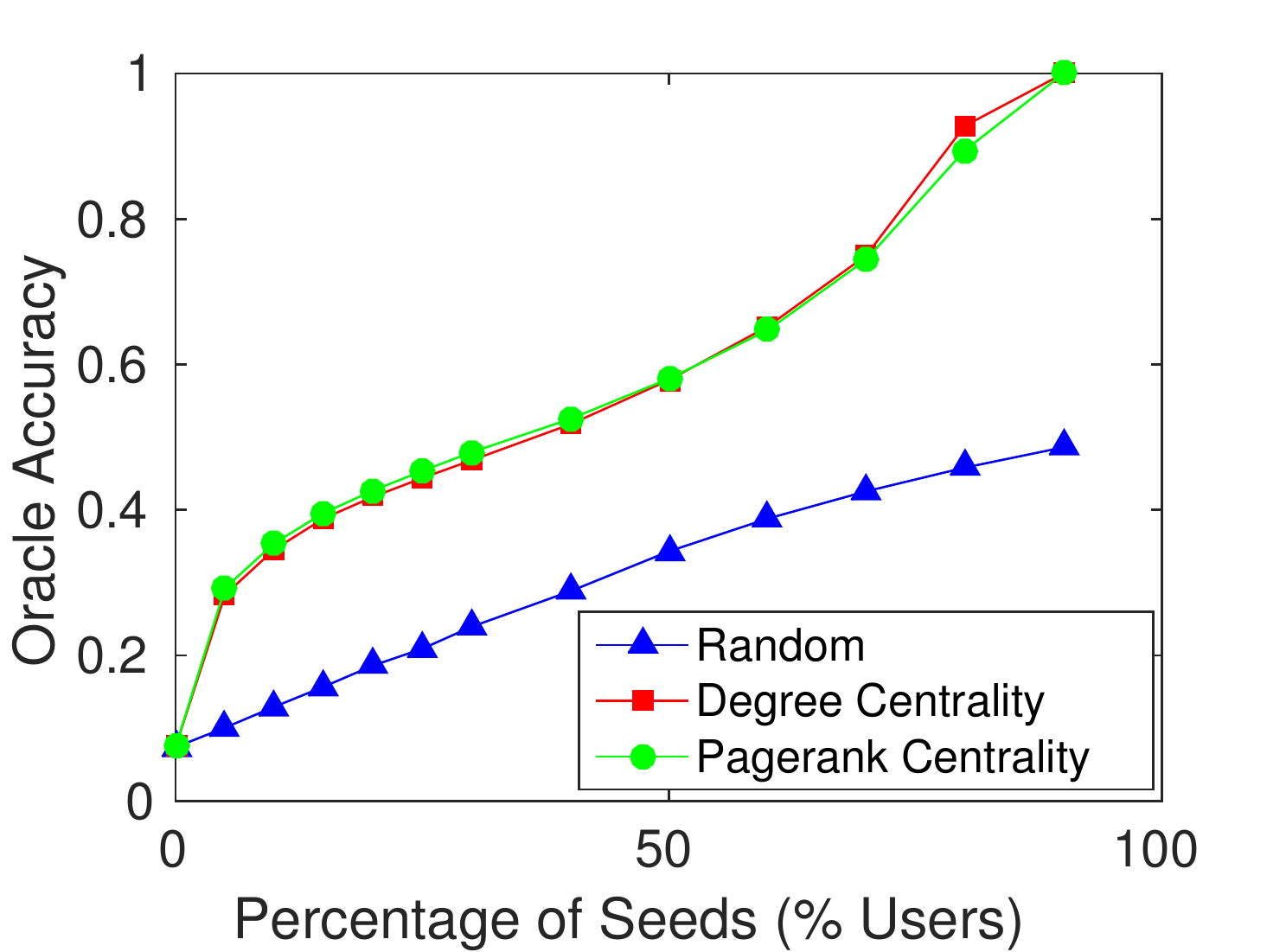}
\caption{$10\%$ shared vertices}
\label{fig:oa10}
\end{subfigure}
\begin{subfigure}{.333\linewidth}
\includegraphics[width=1.05\textwidth]{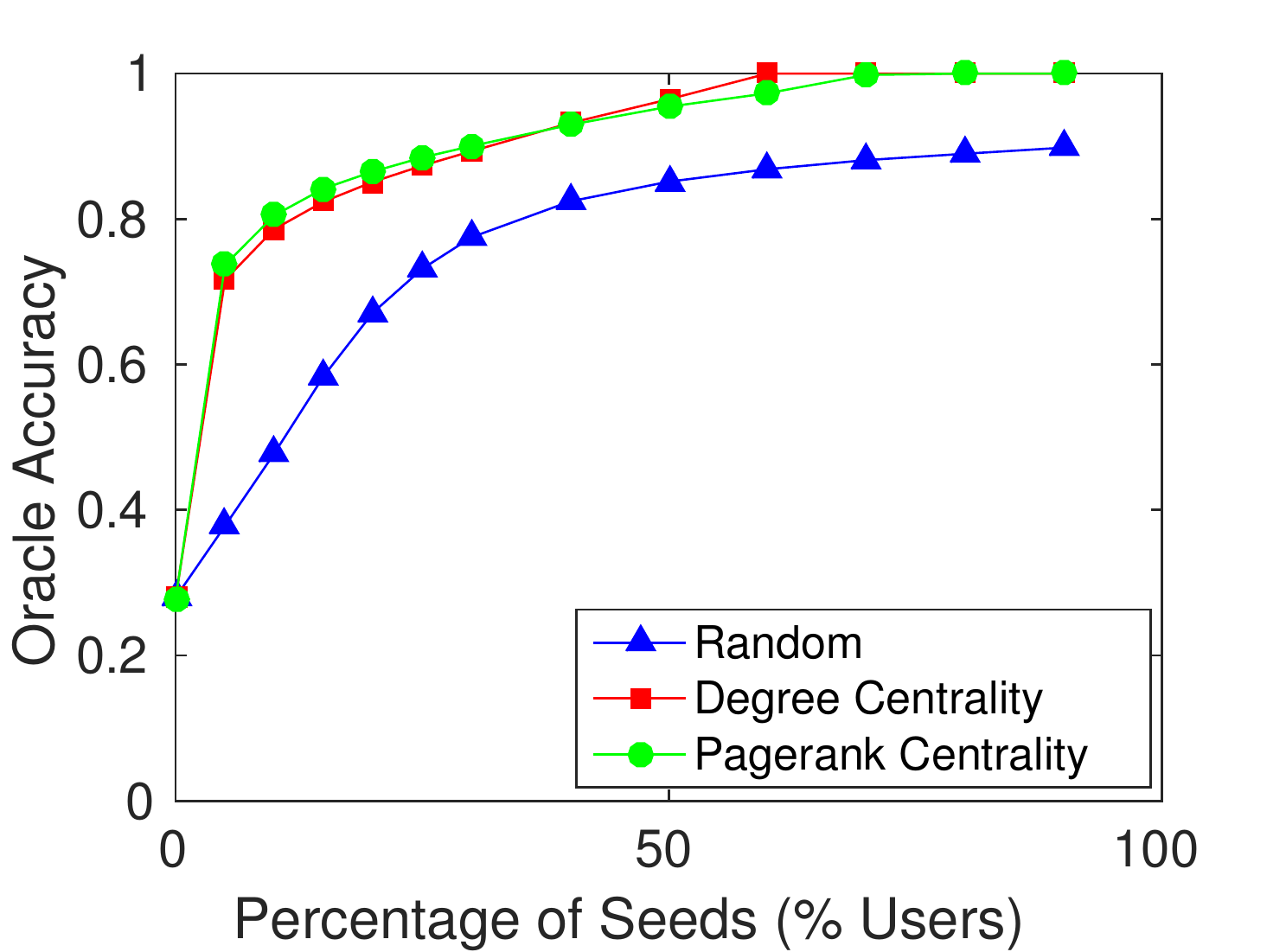}
\caption{$50\%$ shared vertices}
\label{fig:oa50}
\end{subfigure}
\begin{subfigure}{.333\linewidth}
\centering
\includegraphics[width=1.05\textwidth]{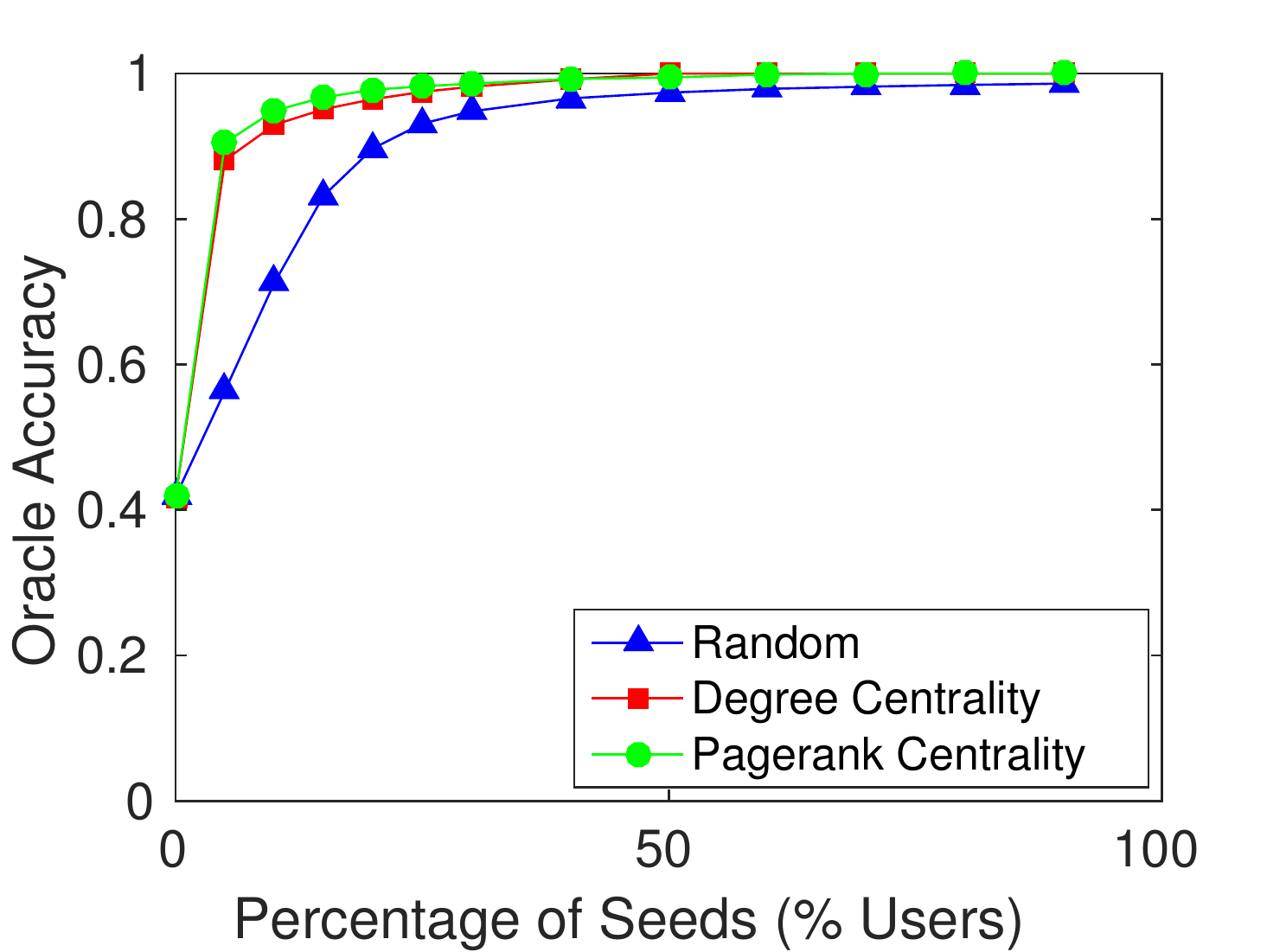}
\caption{$90\%$ shared vertices}
\label{fig:0a90}
\end{subfigure}
\caption{The oracle accuracy results as a function of the percentage of seeds}
\label{fig:oa}
\end{figure}

The plots in Fig.~\ref{fig:oa} show the results of the oracle accuracy as a 
function of the percentage of seeds using the three seed selection strategies. 
Unlike the previous two measures which compare the detected community 
assignment 
$C'$ with the reference community assignment $C$, the oracle accuracy measures 
the percentage of shared users across networks that are in the same 
community; 
we exclude the seed pairs in the computation. As shown in the plots in Fig.~\ref{fig:oa}, both the PageRank and the degree strategies perform significantly 
better than random seed selection. Moreover, from \ref{fig:oa}(b) and 
 \ref{fig:oa}(c), we observe that for networks with large fraction of shared 
users, one only needs a small percentage of seeds to achieve a good oracle 
accuracy. On the other hand, it can be seen from Fig.~\ref{fig:oa}(c) that for 
networks with a small fraction of shared vertices, one needs a large percentage 
of 
seeds to achieve the same level of oracle accuracy. 

\subsection{Twitter-Instagram Networks}\label{sec:ti_exp}
%
%
We now turn to another experiment that explores the possibility of detecting 
communities across Twitter and Instagram networks. 
In this case, we do not have the complete information of the user alignment 
$\mathcal{T}$. However, there are several ways that one can extract some alignment from the data, i.e., username matching, profile information and Instagram 
repost on Twitter. Out of 522,247 Twitter users and 1,045,617 Instagram 
users, we have extracted 42,703 pairs of usernames across the networks that 
are likely to correspond to the same person. We split up these pairs of 
usernames into two sets: training and evaluation. The training set is 
used as seeds $\mathcal{S}$ for community detection, while the evaluation set 
is used to 
compute the oracle accuracy of the detected community structure. The 
aggregation approach is used for combining the Twitter-Instagram networks. The 
adjacency matrices of the individual graphs are normalized on the left by the weighted vertex 
degree.

Table~\ref{table:twitter-instagram} displays the oracle accuracy results of 
different seed selection strategies with varying number of seeds  used in the 
detection. Given that we do not know how many users are shared in both 
networks, the percentage of seeds shown in the table are computed with respect 
to the total number of Twitter users. For the degree and PageRank strategies, 
we first compute the respective centrality values in Twitter and Instagram networks 
separately, and then take the average for each pair of seeds. The result in 
Table~\ref{table:twitter-instagram} shows that in general, the PageRank 
strategy for seed selection gives the best oracle accuracy, suggesting that for 
applications where aligning users across network is costly and there is limited 
resource to cover the cost, it is better to map users who have high PageRank 
centrality as compared with the other two strategies.

\begin{table}[hbt!]
\begin{center}
\caption{The oracle accuracy results of three different seed selection 
strategies on Twitter-Instagram networks}\label{table:twitter-instagram}
\begin{tabular}{l*{6}{c}} 
Seed Selection             & $0.8\%$ & $2.5\%$ & $4\%$ & $5.7\%$ & $7.3\%$ 
&$8\%$\\
\hline
Random 	& 7.72 & 10.62 & 13.37 & 15.62 &17.66 & 23.47  \\
Degree    & 10.93 & 13.62 & 14.81 & 16.51 & 20.10 &19.67  \\
PageRank    & 11.55 & 14.33 & 15.74 & 17.31 & 22.06 & 44.26   \\
\end{tabular}
\end{center}
\end{table}

It is important to note that the seed strategies displayed in 
Table~\ref{table:twitter-instagram} are performed on the known set of aligned users, instead of on the entire set of shared users. Hence, the actual 
oracle accuracy results using the true seed selection strategies should be 
higher than the displayed values. Nonetheless, they should follow a similar 
trend as shown in the table. In addition, for the purpose of this paper, seeds 
are extracted using very basic strategies, such as exact username matching and 
information provided in user profiles and URL links. Hence, there are 
errors in the seed set which may have negatively affected the results. For the 
future work, we will explore more sophisticated methods for obtaining the seed 
set. It is believed that a high quality seed set will significantly improve the 
results.

\section{Conclusion}\label{sec:conclusion}
%
%
In this paper, we explored several methods for community detection across 
online social networks with no prior knowledge of the alignment of users and of 
the fraction of shared users across networks. Specifically, we examined three 
multilayer community detection methods --- aggregation, linking and relaxed 
random walk. We showed that for our problem setup, the aggregation approach 
finds a community assignment that is the most faithful to the underlying 
community structure.
In addition, we examined the effect of seeds 
(which are the extracted mappings of users across networks) on the performance 
of community detection. We proposed three measures to assess the 
performance: variation of information, Jaccard similarity matrix and oracle 
accuracy. We showed that for networks with a large fraction of shared users, 
a small percentage of seeds is needed to obtain a community 
structure that is close to the reference community assignment and has good 
oracle accuracy. However, when the networks have a small fraction of shared 
users, it is difficult to find a good community assignment. We also 
explored three seed selection strategies: random, degree and PageRank. Both the 
degree and PageRank strategies significantly outperformed random seed selection, 
and the PageRank strategy had the best performance. 

For future work, we note that both the quantity and the 
quality of seeds are extremely important in obtaining a good community 
structure across networks. Although we examined several seed selection 
strategies, it is also important to explore methods for (high accuracy) seed 
set expansion, particularly by combining the state-of-the-art entity resolution 
techniques with network structures.  

\bibliographystyle{plain} \small
\bibliography{CommDetect}
\end{document}